\documentclass[%
 reprint,
 amsmath,amssymb,
 aps,
]{revtex4-1}

\usepackage[utf8]{inputenc}
\usepackage{graphicx}
\usepackage[labelformat=simple]{subcaption}
\usepackage{miller}
\usepackage{siunitx}
\usepackage[british]{babel}
\usepackage{hyperref}
\usepackage{ifthen}
\usepackage{booktabs}
\usepackage{xfrac} 
\usepackage{paralist} 
\usepackage{mhchem} 

\newcommand{\paperTitle}{A First Principles Study of Zirconium Grain Boundaries}

        {\end{todosublist}}
\newenvironment{todosublist}{%
    \begin{itemize}%
        \color{Bittersweet}%
        \setlength{\itemsep}{1pt}%
              \setlength{\parskip}{0pt}%
              \setlength{\parsep}{0pt}%
              }
              {\end{itemize}}

\newcommand{\includesvgformatted}[2][]{%
    \ifthenelse{\equal{#1}{}}{%
        {\footnotesize
                {\sffamily
                    {\sansmath
                        \includesvg{#2}}}}%
    }{%
        {\footnotesize
                {\sffamily
                    {\sansmath
                        \includesvg[width=#1]{#2}}}}%
    }%
}

\renewcommand{\vec}[1]{\mathbf{#1}}
\newcommand{\tn}{\mathrm}
\newcommand{\sig}[1]{\ensuremath{\Sigma #1}}
\makeatletter
\newcommand\myIfEmptyTF[1]
{%
    \if\relax\detokenize{#1}\relax
        \expandafter\@firstoftwo
    \else
        \expandafter\@secondoftwo
    \fi
}
\makeatother

\newcommand{\gbWidth}{\ensuremath{w}}
\newcommand{\natom}[1]{\ensuremath{n_\tn{#1}}}
\newcommand{\wSep}[2]{\ensuremath{W_\tn{#1}\defConfig{#2}}}
\newcommand{\eTot}[2]{\ensuremath{E^\tn{tot}_\tn{#1}\defConfig{#2}}}
\newcommand{\en}[1]{\ensuremath{E_\tn{#1}}}
\newcommand{\latAExp}{\ensuremath{a_\tn{exp}}}
\newcommand{\latCExp}{\ensuremath{c_\tn{exp}}}
\newcommand{\latADft}{\ensuremath{a}}
\newcommand{\latCDft}{\ensuremath{c}}
\newcommand{\misori}{\ensuremath{\theta}}
\newcommand{\interplanarSpace}[2]{\myIfEmptyTF{#1}%
    {\ensuremath{d_{ij}}}{\ensuremath{d_{#1#2}}}%
}
\newcommand{\interfaceDist}{\ensuremath{d}}
\newcommand{\coordNum}{\ensuremath{n_\tn{c}}}
\newcommand{\gbExpSym}{\ensuremath{\Delta L}}
\newcommand{\gbExpMinSym}{\ensuremath{\left(\gbExpSym{}\right)_\tn{min}}}
\newcommand{\gbTransSym}{\ensuremath{\vec{t}}}
\newcommand{\gbTransMinSym}{\ensuremath{\gbTransSym{}_\tn{min}}}
\newcommand{\atomVolZr}{\ensuremath{v_\tn{Zr}}}
\newcommand{\atomVolChangeZr}{\ensuremath{\delta v}}

\newcommand{\atomVolPostRelax}{\ensuremath{v}}
\newcommand{\defConfig}[1]{\ensuremath{\myIfEmptyTF{#1}{}{\left[#1\right]}}}
\newcommand{\nCSL}{\ensuremath{n_\tn{CSL}}}
\newcommand{\supVol}{\ensuremath{V_\tn{sup}}}
\newcommand{\supArea}{\ensuremath{A_\tn{sup }}}
\newcommand{\supVec}[1]{%
    \ensuremath{%
        \ifthenelse{\equal{#1}{0}}{\vec{a}}{}%
        \ifthenelse{\equal{#1}{1}}{\vec{b}}{}%
        \ifthenelse{\equal{#1}{2}}{\vec{c}}{}%
    }%
}

\newcommand{\latticeIndexOne}{I}
\newcommand{\latticeIndexTwo}{II}

\newcommand{\alphaZr}{$\alpha$-Zr}
\newcommand{\gammaSurf}{$\gamma$-surface}
\newcommand{\castepVers}{17.2}
\newcommand{\lammpsVers}{11 August, 2017} 
\newcommand{\zrPseudoPotStates}{\ce{4s^2}, \ce{4p^6}, \ce{4d^2} and \ce{5s^2}}

\newcommand{\latAExpDefn}{$\latAExp{}=\angs{3.2316}$}
\newcommand{\latCExpDefn}{$\latCExp{}=\angs{5.1475}$}
\newcommand{\latADftDefn}{$\latADft{}=\angs{3.2301}$}
\newcommand{\latCDftDefn}{$\latCDft{}=\angs{5.1641}$}
\newcommand{\convergeEGbSize}{\jpermsq{0.02}}
\newcommand{\smearingWidthVal}{\evolt{0.1}}
\newcommand{\convergeEPerAtomCutOff}{\evperatom{0.01}}
\newcommand{\cutOffEnergyVal}{\evolt{360}}
\newcommand{\cutOffEnergyValEq}{\ensuremath{\cutOffEnergySym{}=\cutOffEnergyVal{}}}
\newcommand{\cutOffEnergySym}{\ensuremath{E_\tn{cut}}}
\newcommand{\scfDETol}{\evperatom{1e-7}}
\newcommand{\geomDETol}{\evperatom{1e-6}}
\newcommand{\geomFMaxTol}{\evperang{1e-2}}
\newcommand{\convergeEGbKpoints}{\jpermsq{0.02}}
\newcommand{\kpointVal}{\perangs{0.04}}

\newcommand{\eGbFsEq}{%
    \en{GB/FS} = \frac{1}{2\supArea{}} \left(%
    \eTot{GB/FS}{} - \frac{\natom{GB/FS}}{\natom{B}} \eTot{B}{}%
    \right)%
}
\newcommand{\wSepEq}{%
    \wSep{GB/B}{} = \frac{1}{2\supArea{}} \left( 2\eTot{FS}{} - \eTot{GB/B}{} \right)%
}

\newcommand{\jpermsq}[1]{\SI[]{#1}{\joule\per\meter\squared}}
\newcommand{\evperang}[1]{\SI[]{#1}{\electronvolt\per\angstrom}}
\newcommand{\evperangsq}[1]{\SI[]{#1}{\electronvolt\per\angstrom\squared}}
\newcommand{\evperatom}[1]{\SI[]{#1}{\electronvolt.{atom}^{-1}}}
\newcommand{\evolt}[1]{\SI[]{#1}{\electronvolt}}
\newcommand{\angs}[1]{\SI[]{#1}{\angstrom}}
\newcommand{\squareangs}[1]{\SI[]{#1}{\angstrom\squared}}
\newcommand{\perangs}[1]{\SI[]{#1}{\per\angstrom}}
\newcommand{\degs}[1]{\SI[]{#1}{\degree}}
\newcommand{\percent}[1]{\SI[]{#1}{\percent}}
\newcommand{\kelvin}[1]{\SI{0}{\kelvin}}

\newcommand{\gbPlane}[2]{%
    \ifthenelse{\equal{#1}{s7}}{\ifthenelse{\equal{#2}{tlA}}{\hkl(1 -3 2 0)}{}}{}%
    \ifthenelse{\equal{#1}{s7}}{\ifthenelse{\equal{#2}{tlB}}{\hkl(4 -5 1 0)}{}}{}%
    \ifthenelse{\equal{#1}{s13}}{\ifthenelse{\equal{#2}{tlA}}{\hkl(1 -4 3 0)}{}}{}%
    \ifthenelse{\equal{#1}{s19}}{\ifthenelse{\equal{#2}{tlA}}{\hkl(3 -5 2 0)}{}}{}%
    \ifthenelse{\equal{#1}{s31}}{\ifthenelse{\equal{#2}{tlA}}{\hkl(1 -6 5 0)}{}}{}%
    \ifthenelse{\equal{#1}{s7}}{\ifthenelse{\equal{#2}{tw}}{\hkl(0 0 0 1)}{}}{}%
    \ifthenelse{\equal{#1}{s13}}{\ifthenelse{\equal{#2}{tw}}{\hkl(0 0 0 1)}{}}{}%
    \ifthenelse{\equal{#1}{s19}}{\ifthenelse{\equal{#2}{tw}}{\hkl(0 0 0 1)}{}}{}%
}
\newcommand{\gbRotAng}[1]{%
    \ifthenelse{\equal{#1}{s7}}{\degs{21.79}}{}%
    \ifthenelse{\equal{#1}{s13}}{\degs{27.80}}{}%
    \ifthenelse{\equal{#1}{s19}}{\degs{13.17}}{}%
    \ifthenelse{\equal{#1}{s31}}{\degs{17.90}}{}%
}
\newcommand{\gbNumCSL}[2]{%
    \ifthenelse{\equal{#2}{tlA}}{\ifthenelse{\equal{#1}{s7}}{6}{}}{}%
    \ifthenelse{\equal{#2}{tlA}}{\ifthenelse{\equal{#1}{s13}}{4}{}}{}%
    \ifthenelse{\equal{#2}{tlA}}{\ifthenelse{\equal{#1}{s19}}{4}{}}{}%
    \ifthenelse{\equal{#2}{tlA}}{\ifthenelse{\equal{#1}{s31}}{4}{}}{}%
    \ifthenelse{\equal{#2}{tlB}}{\ifthenelse{\equal{#1}{s7}}{10}{}}{}%
    \ifthenelse{\equal{#2}{tw}}{6}{}%
}
\newcommand{\gbNumAtom}[2]{%
    \ifthenelse{\equal{#1}{s7}}{\ifthenelse{\equal{#2}{tlA}}{84}{}}{}%
    \ifthenelse{\equal{#1}{s13}}{\ifthenelse{\equal{#2}{tlA}}{104}{}}{}%
    \ifthenelse{\equal{#1}{s19}}{\ifthenelse{\equal{#2}{tlA}}{152}{}}{}%
    \ifthenelse{\equal{#1}{s31}}{\ifthenelse{\equal{#2}{tlA}}{248}{}}{}%
    \ifthenelse{\equal{#1}{s7}}{\ifthenelse{\equal{#2}{tlB}}{140}{}}{}%
    \ifthenelse{\equal{#1}{s7}}{\ifthenelse{\equal{#2}{tw}}{84}{}}{}%
    \ifthenelse{\equal{#1}{s13}}{\ifthenelse{\equal{#2}{tw}}{156}{}}{}%
    \ifthenelse{\equal{#1}{s19}}{\ifthenelse{\equal{#2}{tw}}{228}{}}{}%
}
\newcommand{\gbTranslation}[2]{%
    \ifthenelse{\equal{#1}{s7}}{\ifthenelse{\equal{#2}{tlA}}{ (\sfrac{1}{4}, \sfrac{1}{2}) }{}}{}%
    \ifthenelse{\equal{#1}{s13}}{\ifthenelse{\equal{#2}{tlA}}{ (\sfrac{1}{2}, \sfrac{1}{2}) }{}}{}%
    \ifthenelse{\equal{#1}{s19}}{\ifthenelse{\equal{#2}{tlA}}{ (0, \sfrac{1}{5}) }{}}{}%
    \ifthenelse{\equal{#1}{s31}}{\ifthenelse{\equal{#2}{tlA}}{ (\sfrac{1}{2}, \sfrac{1}{2}) }{}}{}%
    \ifthenelse{\equal{#1}{s7}}{\ifthenelse{\equal{#2}{tlB}}{ (0, \sfrac{2}{5}) }{}}{}%
    \ifthenelse{\equal{#1}{s7}}{\ifthenelse{\equal{#2}{tw}}{ (0, 0) }{}}{}%
    \ifthenelse{\equal{#1}{s13}}{\ifthenelse{\equal{#2}{tw}}{ (0, 0) }{}}{}%
    \ifthenelse{\equal{#1}{s19}}{\ifthenelse{\equal{#2}{tw}}{ (0, 0) }{}}{}%
}
\newcommand{\gbExpansion}[2]{%
    \ifthenelse{\equal{#1}{s7}}{\ifthenelse{\equal{#2}{tlA}}{0.18}{}}{}%
    \ifthenelse{\equal{#1}{s13}}{\ifthenelse{\equal{#2}{tlA}}{0.16}{}}{}%
    \ifthenelse{\equal{#1}{s19}}{\ifthenelse{\equal{#2}{tlA}}{0.17}{}}{}%
    \ifthenelse{\equal{#1}{s31}}{\ifthenelse{\equal{#2}{tlA}}{0.21}{}}{}%
    \ifthenelse{\equal{#1}{s7}}{\ifthenelse{\equal{#2}{tlB}}{0.24}{}}{}%
    \ifthenelse{\equal{#1}{s7}}{\ifthenelse{\equal{#2}{tw}}{0.10}{}}{}%
    \ifthenelse{\equal{#1}{s13}}{\ifthenelse{\equal{#2}{tw}}{0.09}{}}{}%
    \ifthenelse{\equal{#1}{s19}}{\ifthenelse{\equal{#2}{tw}}{0.10}{}}{}%
}

\begin{document}

\title{\paperTitle{}}
\author{A.\,J.\,Plowman}
\author{C.\,P.\,Race}
\affiliation{Department of Materials, University of Manchester}
\date{\today}

\begin{abstract}
    We present the results of first-principles calculations of selected structural and thermodynamic properties of a set of grain boundaries (GBs) in zirconium, spanning a range of misorientation angles and boundary planes. We performed plane-wave density functional theory calculations on low-\sig{} grain boundaries\,---\,five symmetric tilt GBs (STGBs) and three twist GBs; all with misorientation axes about \hkl[0001] and in optimised microscopic configurations\,---\,to gain insight into the associated atomistic structures. From studying the interface energetics, we found that higher GB excess volumes tended to be associated with higher GB energies. Furthermore, we examined how the interplanar spacing, volume per atom, \atomVolPostRelax{}, and local atomic coordination at the GB deviated from equivalent quantities in bulk. We also defined a grain boundary width according to a threshold value of \atomVolPostRelax{}, allowing us to rank the GBs by their relative thickness. We found the twist GBs to exhibit similar energetic and structural properties, whereas the STGBs demonstrated more variation. Our comprehensive analysis demonstrates how all five dimensions of GB space are crucial in determining properties such as the work of ideal separation and the length scale over which atoms are perturbed by the presence of the GB. So that our results can be useful for further investigations, we have published our data to a public repository (Zenodo). This data includes the optimised and initial structures, in addition to the computed interface energetics and structural properties.
\end{abstract}

\maketitle

\section{Introduction}

Grain boundaries are paramount in determining macroscopic materials properties in polycrystalline materials. For instance, in polycrystalline metals, grain refinement, in which grain size is controlled, can be used to improve the strength of a material \cite{MurtyGrainrefinementaluminium2002}. In order to optimise the performance of engineering materials, and thereby improve their capabilities, an understanding of the processes that limit their suitability must be developed. In many polycrystalline materials, including Zr, GBs play an important role in ultimately determining component strength. For example, intergranular fracture is an observed failure mode of zirconium alloy cladding in light water nuclear reactors \cite{PiroReviewPelletClad2017}. In particular, a form of stress corrosion cracking is believed to originate at the cladding inner surface due to the aggressive action of fission products such as iodine, and the hoop stresses that develop on the cladding as the fuel rod is `burnt'. This so-called pellet-cladding interaction is an example of a complicated process that is difficult to mechanistically untangle from experimental observations alone. First-principles calculations enable an accurate examination of some key aspects of material failure. Crucially, with electronic structure methods such as density functional theory (DFT), which is employed in this work, analyses of atomistic structures are not biassed by transferability issues \cite{KarlsTransferabilityEmpiricalPotentials2016} associated with the use of empirical potentials. In other words, mechanisms operating at the atomistic level can be more reliably probed since the influence of the electronic structure is explicitly included.

Despite the importance of Zr alloys in nuclear reactor applications, there has not been a great deal of research into understanding Zr GB properties, especially not with the accuracy provided by first principles methods. This is partly due to the difficulties associated with constructing a representative set of GB models in Zr, which in the $\alpha$ phase, adopts a hexagonal close-packed structure. As part of their work studying defect segregation, Christensen et al.\ \cite{ChristensenEffectimpurityalloying2010} examined with density functional theory (DFT) a single \sig{7} twist GB about \hkl[0001] and determined the grain boundary energy (\jpermsq{0.29}) and work of ideal separation (the energy difference associated with separating a GB into two free surfaces) of both the GB (\jpermsq{2.86}) and bulk at the same cleavage plane (\jpermsq{3.15}).

Existing work in which multiple GB planes have been probed included that of Uesugi and Higashi \cite{UesugiFirstprinciplescalculationgrain2011}, who performed first principles calculations of six STGBs in Al about the \hkl[110] axis, with the aim of understanding the role of elastic energy in GBs. In addition to calculating the GB energy, the researchers also examined, for each system, the GB excess volume (or GB expansion), which is the local expansion of the crystals at the GB that is typically necessary to accommodate the lattice mismatch. They found a linear relationship between GB energy and excess volume; the computed excess volumes of their studied system ranged from approximately \angs{0.05} to \angs{0.3}; these correlated positively with the GB energies, which were in the range from approximately \jpermsq{0.1} to \jpermsq{0.5}.

Guhl et al.\ \cite{GuhlStructuralelectronicproperties2015} investigated the structural and electronic properties of two STGBs about \hkl[0001] in \ce{Al_2O_3}. They used simulated annealing under an empirical potential to generate a set of configurations for each GB that were thermodynamically reasonable. They then used DFT to examine the properties of each GB, including the local atomic coordination; four-fold \ce{Al} sites were found to be associated with lower energy GB configurations. General agreement was found between the results of their employed empirical potential and those of DFT, although the authors discussed some exceptions to this.

For the first time, we have probed from first-principles, a set of multiple GBs in Zr that includes multiple GB planes. We have studied the first four low-\sig{} GBs (\sig{7}, \sig{13}, \sig{19}, and \sig{31}) in Zr with misorientation axes about [0001]. The constructed systems consist of five symmetric tilt GBs (STGBs, including two distinct STGBs with the same \sig{}-value), and three twist GBs. We have examined the interfacial energetics and focussed on key atomistic structural properties, such as coordination and volume per atom. We have demonstrated that the GB plane is just as important in determining GB properties as are the misorientation axis and angle.

\section{Method}

\subsection{Atomistic models}

Grain boundary simulation cells (supercells) were constructed using the coincident site lattice (CSL). At particular misorientation angles between two superimposed identical lattices, say, lattice \latticeIndexOne{} and lattice \latticeIndexTwo{}, some proportion (conventionally denoted $1/\sig{}$) of lattice sites precisely coincide, enabling the construction of a CSL unit cell. Periodic GB supercells may be formed by adjoining two CSL unit cells, where one is populated with lattice \latticeIndexOne{} sites and the other with lattice \latticeIndexTwo{} sites. A valid CSL unit cell may be formed by choosing as its edge vectors any linear combination of CSL \emph{primitive} unit cell edge vectors. In this way, supercells containing various GB planes may be constructed.

A GB model constructed via the CSL\footnote{or, indeed, a GB in a real material that approximately resembles a CSL GB.} may be macroscopically denoted as follows: \sig{p}\hkl(hkil)\hkl[uvtw], where $p$ is the \sig{}-value (i.e.\ $\sig{}=p$) associated with forming a CSL due to a relative rotation between lattices \latticeIndexOne{} and \latticeIndexTwo{}, about the crystallographic axis \hkl[uvtw]. The GB plane\,---\,which is independent of both the \sig{}-value and the rotation axis\,---\,is specified by \hkl(hkil). For the STGBs and twist GBs studied in this work, this specification is complete; for a general (non-symmetric) GB, two planes would typically be included in the specification.

Owing to the computational expense of first-principles simulation techniques, we may submit only relatively small systems for examination. Consequently, in this study, as in those in the literature \cite{WangGrainboundariesbccFe2018}, we were limited to investigating so-called low-\sig{} GBs, which can be represented within relatively small supercells. Despite this limitation, we found the choice of GB plane to be just as important as the \sig{}-value in determining GB properties such as energetics and local atomic structure, as discussed later. It is therefore reasonable to assume that we have probed a broad set of atomistic environments from amongst those likely to be found in real materials.

In Table \ref{gbp:tab:interfaceConstruction}, we list the eight GBs that were investigated. In particular, we explored two types of boundary planes: five tilt GBs, in which the GB plane is in the zone of the rotation axis, and three twist GBs, in which the GB plane is normal to the rotation axis. We note that all tilt boundaries were also symmetric, meaning the specification of the boundary plane, \hkl(hkil), is symmetrically equivalent in both micro-grain crystal bases. We refer to these boundaries as STGBs. All GBs were formed via rotations about \hkl[0001]. Since, for \alphaZr{}, the square of the ratio of the hexagonal lattice parameters, $(c/a)^2$, is irrational, CSL GBs cannot be formed about arbitrary axes. Although this is somewhat restrictive, we are still able to explore \begin{inparaenum}[i)]\item a large range of misorientation angles, including low- and high-angle GBs, and \item the effects of different GB planes.\end{inparaenum}

In Table \ref{gbp:tab:interfaceConstruction}, we distinguish two types of STGB: type I and type II (not to be confused with our labelling of lattices I and II in the previous discussion of the CSL). When generating STGBs, we may construct the CSL unit cells such that the GB plane in the resulting GB supercell is any crystallographic plane that is in the zone of the rotation axis. However, we must be mindful of computational expense; system sizes should be kept minimal. Thus, those STGBs labelled as `type I' in Table \ref{gbp:tab:interfaceConstruction} are the boundaries that are computationally the cheapest for that \sig{}-value. In particular, their respective CSL unit cells can be considered primitive. On the other hand, the single \sig{7} STGB that is labelled as `type II' is the next-cheapest STGB that can be constructed for this \sig{}-value. To be precise, the CSL unit cells associated with the type I and type II STGBs have the same volume, but the type II unit cell has a smaller perpendicular distance separating the two GB planes, and so more repeats are necessary in order to avoid spurious interactions between the two GBs within the supercell. The \sig{13}, \sig{19} and \sig{31} type I STGBs are also referenced without the `type' label, since it is only the \sig{7} STGB for which multiple `types' were studied.

For each GB supercell, complementary bulk and free surface (FS) supercells were constructed. To generate a FS model, atoms belonging to one micro-grain in the GB model were removed. To then generate a bulk model, the remaining atoms in the FS model were copied into the vacuum region.

\begin{table}[ht]
    \centering
    \caption{Properties of the interface models studied in this work, including GB misorientation angle, \misori{}, total number of CSL unit cells per GB supercell, \nCSL{}, and number of atoms per GB supercell, \natom{GB}. From explorations of the \gammaSurf{}s, the minimum-energy relative translation, \gbTransMinSym{}, and the minimum-energy GB expansion, \gbExpMinSym{} (expressed in \angs{}), were also found. All GBs were formed by rotations about \hkl[0001].}
    \label{gbp:tab:interfaceConstruction}
    \begin{tabular}{rlllrrcc}
        \toprule
        \sig{} & Plane              & Type    & \misori{}      & \nCSL{}             & \natom{GB}           & \gbTransMinSym{}         & \gbExpMinSym{}         \\ \midrule
        7      & \gbPlane{s7}{tlA}  & Tilt I  & \gbRotAng{s7}  & \gbNumCSL{s7}{tlA}  & \gbNumAtom{s7}{tlA}  & \gbTranslation{s7}{tlA}  & \gbExpansion{s7}{tlA}  \\
        13     & \gbPlane{s13}{tlA} & Tilt I  & \gbRotAng{s13} & \gbNumCSL{s13}{tlA} & \gbNumAtom{s13}{tlA} & \gbTranslation{s13}{tlA} & \gbExpansion{s13}{tlA} \\
        19     & \gbPlane{s19}{tlA} & Tilt I  & \gbRotAng{s19} & \gbNumCSL{s19}{tlA} & \gbNumAtom{s19}{tlA} & \gbTranslation{s19}{tlA} & \gbExpansion{s19}{tlA} \\
        31     & \gbPlane{s31}{tlA} & Tilt I  & \gbRotAng{s31} & \gbNumCSL{s31}{tlA} & \gbNumAtom{s31}{tlA} & \gbTranslation{s31}{tlA} & \gbExpansion{s31}{tlA} \\
        7      & \gbPlane{s7}{tlB}  & Tilt II & \gbRotAng{s7}  & \gbNumCSL{s7}{tlB}  & \gbNumAtom{s7}{tlB}  & \gbTranslation{s7}{tlB}  & \gbExpansion{s7}{tlB}  \\
        7      & \gbPlane{s7}{tw}   & Twist   & \gbRotAng{s7}  & \gbNumCSL{s7}{tw}   & \gbNumAtom{s7}{tw}   & \gbTranslation{s7}{tw}   & \gbExpansion{s7}{tw}   \\
        13     & \gbPlane{s13}{tw}  & Twist   & \gbRotAng{s13} & \gbNumCSL{s13}{tw}  & \gbNumAtom{s13}{tw}  & \gbTranslation{s13}{tw}  & \gbExpansion{s13}{tw}  \\
        19     & \gbPlane{s19}{tw}  & Twist   & \gbRotAng{s19} & \gbNumCSL{s19}{tw}  & \gbNumAtom{s19}{tw}  & \gbTranslation{s19}{tw}  & \gbExpansion{s19}{tw}  \\ \bottomrule
    \end{tabular}
\end{table}

Plane-wave DFT codes simulate supercells with implicit periodic boundary conditions. Although well-suited to studying crystalline matter in general, the periodicity of the plane-wave basis set can be problematic when studying defects such as grain boundaries. GB supercells are typically constructed either without vacuum, where the supercell contains two (potentially distinct) GBs, or with vacuum, where the supercell contains one GB and two free surfaces. The advantage of including vacuum is the isolation of the GB of interest; in particular, energetic quantities such as the grain boundary energy can be computed for that precise GB. On the other hand, in the fully-periodic model, if the two GBs are not identical, only an \emph{average} quantity could be determined. However, it is preferable to avoid the introduction of free surfaces into the GB supercell, since they may have a significant effect on the atomic relaxation of the GB. Following the ideas of Guhl et al.\ \cite{GuhlStructuralelectronicproperties2015}, we have constructed fully-periodic GB supercells such that the two GBs present in each supercell are identical.

Crucially, by manipulating the skewness of the supercell, we could ensure this equivalence was maintained whilst exploring the microscopic degrees of freedom of each macroscopically defined GB. Such an exploration is necessary to ensure a thermodynamically reasonable system is simulated. For instance, the macroscopic specification of a GB, which comprises the misorientation axis and angle (via the \sig{}-value, in our case), and the GB plane, does \emph{not} describe the relative positions of atoms across neighbouring grains. Thus, atoms may be in unphysical proximity according to the initial construction of the supercell. A more realistic microscopic configuration can be determined by exploring the energy landscape that results from performing total energy calculations after sequential rigid translations of one micro-grain relative to the other. For each point on this energy landscape, known as a \gammaSurf{}, we also explored changes in energy resulting from introducing small amounts of vacuum at the GBs. Thus, for each GB studied, we identified \begin{inparaenum}[i)]\item the minimum-energy configuration in terms of a translation of one micro-grain relative to the other, \gbTransMinSym{}, and \item the GB expansion\end{inparaenum}, written as:

\begin{equation}
    \gbExpMinSym{} = \frac{1}{2\supArea{}} \left(
    \supVol{} - \natom{GB}\atomVolZr{}
    \right),
\end{equation}

\noindent where \supArea{} is the area of the GB (for a parallelepiped supercell defined by edge vectors \supVec{0}, \supVec{1} and \supVec{2}, where edge vectors \supVec{0} and \supVec{1} lie in the GB plane, $\supArea{}=|\supVec{0}\times\supVec{1}|$), \supVol{} is the total volume of the GB supercell ($\supVol{}=|\supVec{0}\cdot(\supVec{1}\times\supVec{2})|$) in the identified minimum-energy configuration, \natom{GB} is the number of atoms in the GB supercell, and \atomVolZr{} is the volume associated with a single bulk \alphaZr{} atom. \gbExpSym{} is  also known as the GB excess volume (e.g.\ in Ref.\ \cite{BeanOrigindifferencesexcess2016}). The factor of two arises since there are two GBs within the supercell, as previously discussed. In Table \ref{gbp:tab:interfaceConstruction}, we include \gbTransMinSym{} and \gbExpMinSym{} for each GB system. Due to all misorientation axes being about \hkl[0001], for the STGBs, the first component of \gbTransMinSym{} is in the direction of the $c$-axis of the hexagonal \alphaZr{} lattice.

\begin{figure}[ht]
    \hspace*{\fill}%
    \subcaptionbox{\label{gbp:fig:gbSchematicsTiltA}}{\includegraphics[scale=0.7]{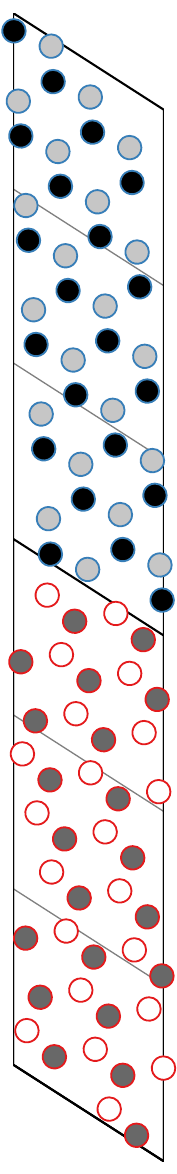}}\hfill%
    \subcaptionbox{\label{gbp:fig:gbSchematicsTiltB}}{\includegraphics[scale=0.7]{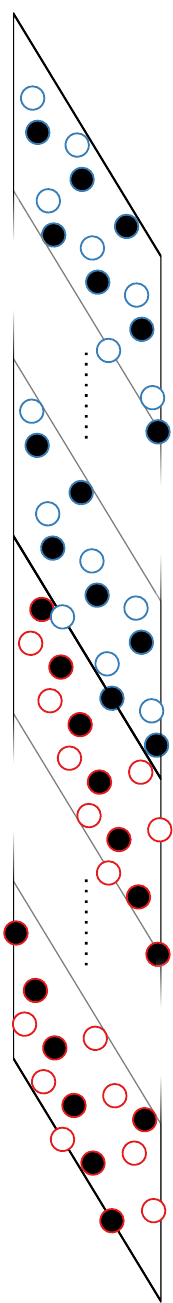}}\hfill%
    \subcaptionbox{\label{gbp:fig:gbSchematicsTiltC}}{\includegraphics[scale=0.7]{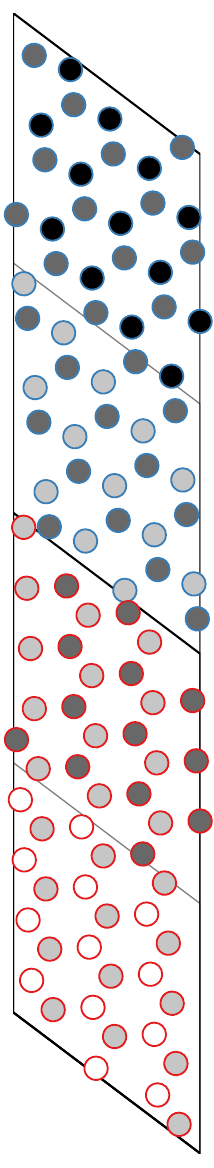}}\hfill%
    \subcaptionbox{\label{gbp:fig:gbSchematicsTiltD}}{\includegraphics[scale=0.7]{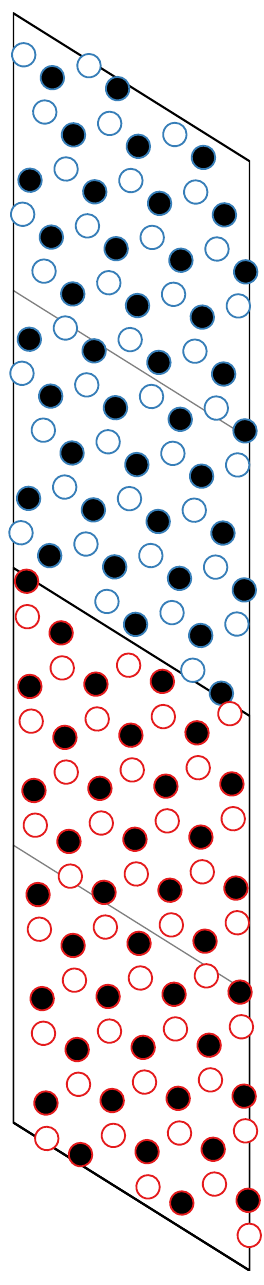}}\hfill%
    \subcaptionbox{\label{gbp:fig:gbSchematicsTiltE}}{\includegraphics[scale=0.7]{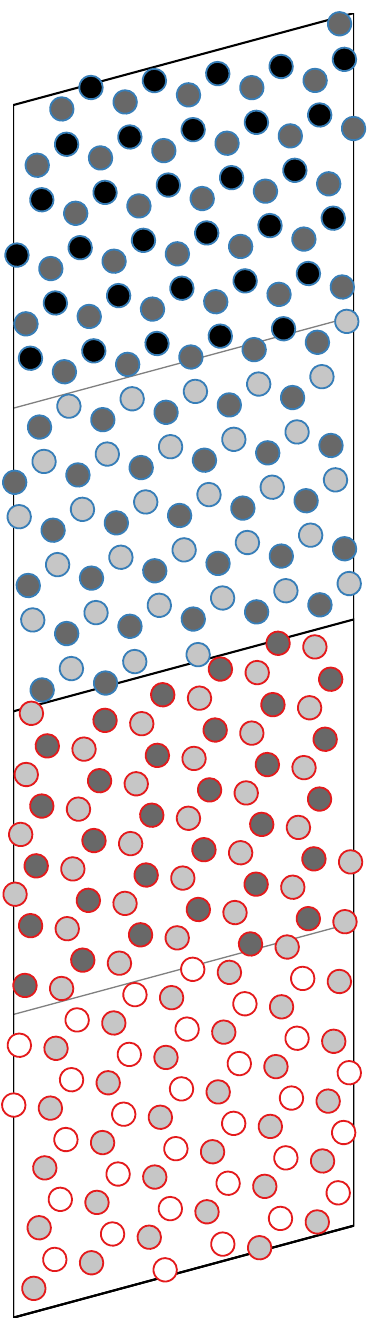}}\hfill%
    \hspace*{\fill}%

    \caption{Schematics of the five STGBs studied in this work: \sig{7} type I STGB \subref{gbp:fig:gbSchematicsTiltA}, \sig{7} type II STGB \subref{gbp:fig:gbSchematicsTiltB}, \sig{13} STGB \subref{gbp:fig:gbSchematicsTiltC}, \sig{19} STGB \subref{gbp:fig:gbSchematicsTiltD} and \sig{31} STGB \subref{gbp:fig:gbSchematicsTiltE}. Periodic supercells are shown and atom sites are illustrated after application of the GB expansions, \gbExpMinSym{}, and relative micro-grain translations, \gbTransMinSym{}, but before any relaxation. Micro-grains are distinguished by atom outline colour (red or blue). CSL unit cells are indicated by grey lines. Atom shade indicates the coordinate in the page-normal direction, where lighter atoms are closer to the reader. The GB rotation axis is parallel to the page-normal direction. All supercell atoms are visible in schematics \subref{gbp:fig:gbSchematicsTiltA} and \subref{gbp:fig:gbSchematicsTiltC}-\subref{gbp:fig:gbSchematicsTiltE}; some atoms have been excluded from \subref{gbp:fig:gbSchematicsTiltB}.}
    \label{gbp:fig:gbSchematicsTilt}
\end{figure}

\begin{figure}[ht]

    \parbox{.45\textwidth}{
        \centering
        \parbox[b]{.2\textwidth}{%
            \hspace*{\fill}\subcaptionbox{\label{gbp:fig:gbSchematicsTwistA}}{\includegraphics[scale=0.7]{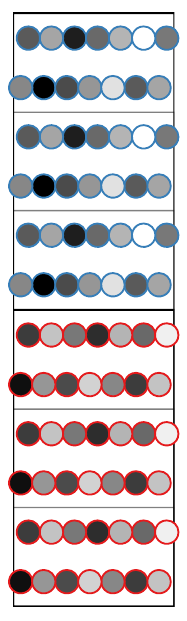}}%
            \vskip1em
            \hspace*{\fill}\subcaptionbox{\label{gbp:fig:gbSchematicsTwistB}}{\includegraphics[scale=0.7]{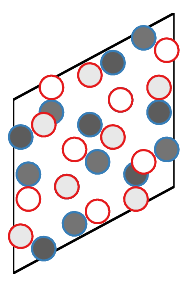}}%
        }
        \hskip1em
        \vspace*{\fill}
        \parbox[b]{.2\textwidth}{%
            \subcaptionbox{\label{gbp:fig:gbSchematicsTwistC}}{\includegraphics[scale=0.7]{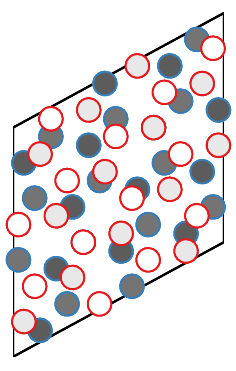}}%
            \vskip1em
            \subcaptionbox{\label{gbp:fig:gbSchematicsTwistD}}{\includegraphics[scale=0.7]{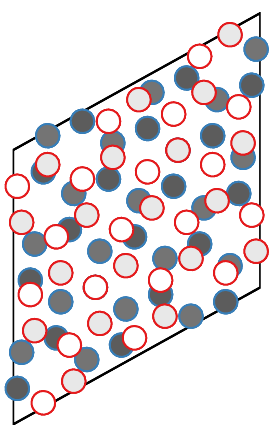}}%
        }
    }
    \caption{Schematics of the three twist GBs studied in this work: \sig{7} twist GB \subref{gbp:fig:gbSchematicsTwistA} and \subref{gbp:fig:gbSchematicsTwistB}, \sig{13} twist GB \subref{gbp:fig:gbSchematicsTwistC}, and \sig{19} twist GB \subref{gbp:fig:gbSchematicsTwistD}. The atomistic configurations are shown after the application of \gbExpMinSym{} and \gbTransMinSym{}, but before geometry optimisation. Micro-grains are distinguished by atom outline colour (red or blue). CSL unit cells are indicated by grey lines. Atom shade indicates the coordinate in the page-normal direction, where lighter atoms are closer to the reader. The GB rotation axis in \subref{gbp:fig:gbSchematicsTwistA} is parallel to the vertical edges of the supercell. The GB rotation axes in \subref{gbp:fig:gbSchematicsTwistB}--\subref{gbp:fig:gbSchematicsTwistD} are parallel to the page-normal direction. The view in \subref{gbp:fig:gbSchematicsTwistB} is formed by looking from the base of the schematic as shown in \subref{gbp:fig:gbSchematicsTwistA} towards the middle GB plane.}
    \label{gbp:fig:gbSchematicsTwist}
\end{figure}

Due to the computational expense of sampling a three-dimensional space with DFT calculations, we did not compute the full \gammaSurf{} for every GB. Instead, we first compared the full \gammaSurf{} of the \sig{7} type I STGB as computed by DFT with that computed by a Zr empirical potential (the embedded-atom method (EAM) potential proposed by Mendelev and Ackland as number 3 in Ref.\ \cite{MendelevDevelopmentinteratomicpotential2007}). Empirical potentials are computationally cheap, but can suffer from transferability issues when simulating `unseen' systems to which they were not fitted \cite{KarlsTransferabilityEmpiricalPotentials2016}. We found the EAM potential to correctly predict the general shape of the \gammaSurf{}, as well as its four-fold symmetry. However, some finer details evident in the DFT \gammaSurf{} were absent when computed with the EAM potential. Nonetheless, the identification of the minimum-energy relative translation could have been made using either the EAM potential or DFT. Consequently, for the larger GBs, we did not need to explore the full \gammaSurf{}. For the \sig{13} STGB, we computed with DFT one quarter of the full \gammaSurf{}\,---\,the symmetrically irreducible region as predicted by the EAM potential. As for the \sig{7} type I STGB, DFT and the EAM potential both predicted an identical minimum-energy relative translation, when considering the resolution with which we sampled the \gammaSurf{} using DFT (for the STGBs, the \gammaSurf{}s were computed on an orthogonal grid of points separated by approximately \angs{1.2} in each orthogonal direction). Due to the observed general agreement between DFT and the EAM potential in the \sig{7} and \sig{13} type I STGBs, we relied solely on the EAM potential to identify minimum-energy relative translations, \gbTransMinSym{}, for the \sig{19}, \sig{31} and \sig{7} type II STGBs. In all cases, the minimum-energy GB expansion, \gbExpMinSym{}, was computed with DFT.

To further reduce the computational expense of exploring the microscopic degrees of freedom of each GB, we imposed constraints on the atoms during \gammaSurf{} and GB expansion calculations; atoms were free to relax only in the GB-normal direction. Once the preferred microscopic configuration was identified for each GB, the supercell was reconstructed in this configuration and atoms were fully relaxed. In all geometry optimisations, the supercell dimensions were fixed.

Upon inspection of the twist GB \gammaSurf{}s as predicted by the EAM potential, the symmetrically irreducible region was found to have a fractional extent of 1/\sig{} in each \gammaSurf{} direction. For example, the symmetrically irreducible region on the \sig{7} twist GB \gammaSurf{} had the same shape as the full \gammaSurf{}, but spanned only 1/49 of its area. This result can be deduced using geometrical arguments, where it can be seen that translating one micro-grain to points on a \sig{}-by-\sig{} grid results in the same relative configuration of atoms across the GB; the only difference is the position of atoms from \emph{both} micro-grains relative to the boundaries of the supercell, which is immaterial (as long as, in the case of a DFT simulation, k-point sampling is sufficiently dense). Energy differences resulting from relative translations within the symmetrically irreducible regions of the twist \gammaSurf{}s were found to be small, as reported by Christensen et al.\ \cite{ChristensenEffectimpurityalloying2010} for the case of the \sig{7} twist GB. Consequently, no relative translations were employed for the twist GBs in this work.

Schematics of the GB systems investigated in this work, after the application of the minimum-energy expansion, \gbExpMinSym{}, and the minimum-energy relative micro-grain translation, \gbTransMinSym{}, but before relaxation of the atomic coordinates, are shown for the STGBs in Fig.\ \ref{gbp:fig:gbSchematicsTilt}, and for the twist GBs in Fig.\ \ref{gbp:fig:gbSchematicsTwist}. In Fig.\ \ref{gbp:fig:gbSchematicsTiltB}, three CSL unit cells have been removed from each micro-grain of the \sig{7} type II STGB schematic and replaced with dashed lines. Note that the supercells in Figs.\ \ref{gbp:fig:gbSchematicsTiltA}, \subref{gbp:fig:gbSchematicsTiltC} and \subref{gbp:fig:gbSchematicsTiltE} (\sig{7}, \sig{13} and \sig{31} type I STGBs, respectively) are skewed in the page-normal direction (as also indicated by the non-zero first components of \gbTransMinSym{} for these GBs in Table \ref{gbp:tab:interfaceConstruction}).

Grain boundary and free surface energies per unit interface area can be computed according to:

\begin{equation}
    \label{eq:gbPropsEGbFs}
    \eGbFsEq{},
\end{equation}

\noindent where \eTot{GB/FS}{} and \eTot{B}{} are the total supercell energies of GB or FS and bulk system supercells, after relaxation of atomic coordinates, and \natom{GB/FS} and \natom{B} are the number of atoms in the GB or FS and bulk system supercells, respectively.

An estimation of the relative strength of a GB (or a bulk system, at a specified interface plane) can be made by computing the work of ideal separation, \wSep{GB/B}{}, as follows:

\begin{equation}
    \label{gbp:eq:wSep}
    \wSepEq{},
\end{equation}

\noindent which is defined so that a positive \wSep{}{} implies work must be done to cleave the system. In this work, we compute \wSep{GB}{} for all GB systems, and \wSep{B}{} for separation of the corresponding bulk systems\,---\,i.e.\ for cleavage at the same plane as in the GB system.

\subsection{Computational parameters}

We employed two classes of atomistic simulation methods to explore Zr GB properties. Primarily, we used the density functional theory (DFT) method, which, through modelling a material's electronic state at the quantum mechanical level (via the electron density) \cite{HohenbergInhomogeneousElectronGas1964,KohnSelfConsistentEquationsIncluding1965}, can offer highly accurate materials properties at the atomistic scale, such as optimal atomic coordinates. Additionally, and as discussed in the previous section, we also relied on, where reasonable, computationally cheap classical molecular statics simulations, for which we employed the Large-scale Atomic/Molecular Massively Parallel Simulator (LAMMPS) software package \cite{PlimptonFastParallelAlgorithms1995} (version dated \lammpsVers{}). To represent the interatomic interactions within LAMMPS simulations, we chose the EAM Zr potential due to Mendelev and Ackland, as described as potential \#3 in Ref.\ \cite{MendelevDevelopmentinteratomicpotential2007}.

For DFT simulations, we used the CASTEP code \cite{ClarkFirstprinciplesmethods2005}, version \castepVers{}. For computational efficiency, CASTEP relies on pseudopotentials, whereby the complex all-electron Coulomb potential of an atom is replaced with a simpler, effective potential that can be represented numerically with greater efficiency. To do this, the non-valence, core electrons are `frozen', meaning it is assumed such electrons do not contribute to bonding. On the other hand, outer electrons, which are more likely to participate in bonding, are explicitly modelled. In this work, we used CASTEP's on-the-fly functionality for generating a suitable ultrasoft pseudopotential for Zr. In particular, the Zr electron states \zrPseudoPotStates{} were explicitly modelled.

In CASTEP, the electronic wave functions are represented using a periodic plane-wave basis, whose size is determined via the cut-off energy parameter, \cutOffEnergySym{}. Initial cut-off energy convergence tests were performed on bulk \alphaZr{}; we selected \cutOffEnergyValEq{} for all subsequent simulations, since total energies were found to be converged to within \convergeEPerAtomCutOff{}.

To represent the quantum-mechanical effects of electron exchange and correlation (XC), we selected the functional due to Perdew, Burke and Ernzerhof (PBE) \cite{PerdewGeneralizedGradientApproximation1996}, which is a type of generalised gradient approximation (GGA) XC functional. The PBE functional is widely used and generally considered to provide relatively accurate results for bulk and surface properties of metals \cite{FiolhaisSurfaceenergiessimple2005}.

To avoid instabilities associated with charge sloshing, a density mixing (DM) scheme \cite{KresseEfficiencyabinitiototal1996} was employed; in particular, that due to Pulay \cite{PulayConvergenceaccelerationiterative1980}. In this approach, the trial density at each step in the search for the ground state density is formed using contributions from a history of previous trial densities. When simulated with DFT, metallic systems are additionally subject to instabilities associated with bands crossing the Fermi level. To avoid the resulting discontinuous occupancies at the Fermi level, a smearing of bands can be applied. In this work, we used a Gaussian smearing of \smearingWidthVal{}. Self-consistent minimisation of the electron density was terminated when the change in energy fell below \scfDETol{}. Geometry optimisations were terminated when the energy and maximum force reached convergence tolerances of \geomDETol{} and \geomFMaxTol{}, respectively. A low-memory implementation of the Broyden-Fletcher-Goldfarb-Shanno (LBFGS) algorithm \cite{ByrdRepresentationsquasiNewtonmatrices1994} was employed for this purpose.

In opting for GB supercells that contain two identical GBs, rather than one GB and two free surfaces, we were constrained to use an integer number of CSL unit cells in the out-of-boundary direction. We performed convergence tests, in which we computed \en{GB} for increasing supercell sizes, to determine how many CSL unit cells, \nCSL{}\footnote{\nCSL{} is the \emph{total} number of CSL unit cells within the supercell, and is thus always an even number.}, should be used for each GB system. For each GB system, \nCSL{}, which is listed in Table \ref{gbp:tab:interfaceConstruction}, was chosen to ensure convergence of \en{GB} to within \convergeEGbSize{}. Furthermore, we also tested the convergence of \en{GB} with respect to k-point sampling density. The Brillouin zone was uniformly sampled according to a Monkhorst-Pack (MP) grid \cite{MonkhorstSpecialpointsBrillouinzone1976}. When the maximum k-point separation was set to \kpointVal{}, \en{GB} was found to be converged to within \convergeEGbKpoints{}, which we deemed sufficient for our purposes.

Values of \eTot{}{} (like those required in Eq.\ \ref{eq:gbPropsEGbFs} for computing the interface energies) were taken from the estimated \kelvin{0} energies reported by CASTEP. Where applicable, we have converted the units of interfacial energies (such as \en{GB}) from \evperangsq{} to SI units of \jpermsq{}, using the physical constants in Ref.\ \cite{MohrCODATARecommendedValues2012}, which is also used internally by CASTEP.

\subsection{Structural analysis}

By measuring the interplanar spacing, we examined the extent to which each GB can perturb atoms from their bulk-like coordinates, as we move further from the interface. This produces some measure of the relative thickness of each GB. In particular, we considered the deviation of interplanar spacing (relative to that in bulk) in the vicinity of each GB by computing, as a function of distance from the GB, the coordinates of crystallographic planes that are parallel to the GB. Where multiple atoms lay in this plane, we took an arithmetic average. We label the resulting distances as \interplanarSpace{}{}, where $i$ and $j$ refer to adjacent plane indices, increasing with distance from the interface (where $j=i+1$). For all systems, except the \sig{7} type II STGB, the $i=0$ plane refers to the plane in the opposing micro-grain such that \interplanarSpace{0}{1} is the distance between planes from different micro-grains. For the \sig{7} type II STGB supercell, which includes atoms precisely on the nominal GB plane, the $i=0$ plane is the nominal GB plane.

To further examine the atomistic structures associated with the studied GBs, we performed a Voronoi tessellation on all GB supercells, both before and after atomic relaxation. To this end, the supercell was partitioned into convex polyhedra that bound the regions of space associated with each individual atom.
The tessellation is defined such that all points in the Voronoi region associated with a given atom are closer to that atom than any other atom. This is a useful analysis method for a number of reasons. The Voronoi tessellation results in an association of each atom with its nearest neighbours (by considering to which pair of atoms each Voronoi facet corresponds). Hence, we can obtain a coordination number for each atom. In doing this, we found it necessary to filter out facets below a certain threshold area, some of which are numerical artefacts due to the degenerate nature of Voronoi vertices in periodic systems. Additionally, we assigned a scalar to each atom: the volume of the Voronoi region, which is helpful when comparing any given atom with the bulk-like environment. Furthermore, we derived a simple measure of the effective GB width, \gbWidth{}, by considering how the volume per atom (relative to that of bulk) changes with distance from the GB plane.

\section{Results and discussion}

\subsection{Bulk \texorpdfstring{\alphaZr{}}{α-Zr} lattice parameters}

We began by performing a geometry optimisation of the two-atom bulk \alphaZr{} primitive unit cell, in order to ascertain the Zr lattice parameters according to the PBE exchange correlation functional. We applied symmetry constraints on the unit cell, such that the hexagonal close-packed structure was maintained during relaxation. Consequently, the optimised lattice parameters were found to be: \latADftDefn{} and \latCDftDefn{}, which agree well with the experimentally determined values of \latAExpDefn{} and \latCExpDefn{} \cite{HaynesCRCHandbookChemistry2014}. All subsequent supercells were constructed using the PBE-optimised lattice parameters.

\subsection{Interface energetics}

\begin{figure}[ht]
    \subcaptionbox{\label{gbp:fig:pristPropsA}}{\includegraphics[scale=0.8]{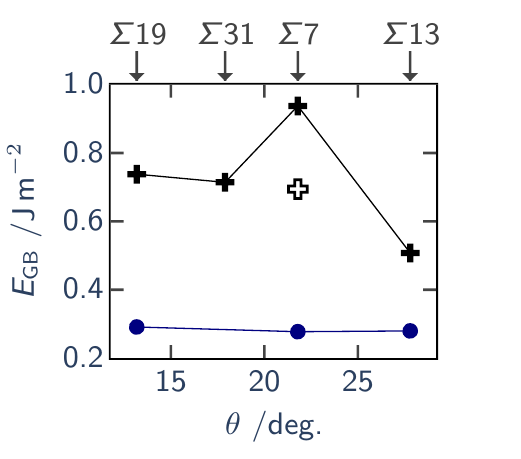}}%
    \subcaptionbox{\label{gbp:fig:pristPropsB}}{\includegraphics[scale=0.8]{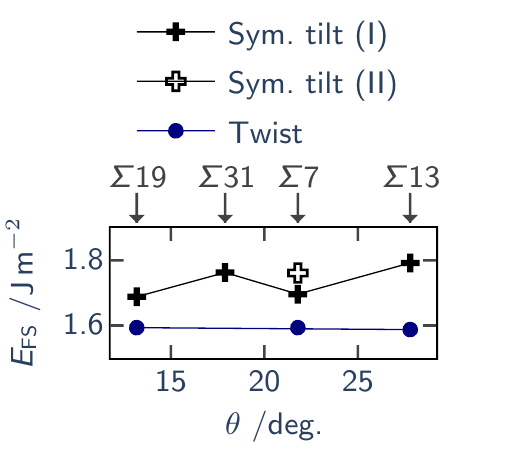}}%

    \bigskip

    \subcaptionbox{\label{gbp:fig:pristPropsC}}{\includegraphics[scale=0.8]{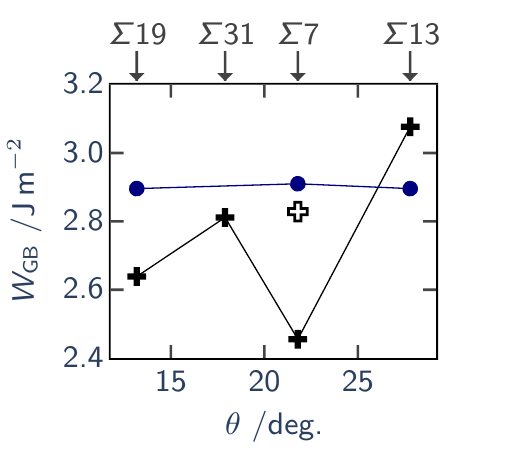}}%
    \subcaptionbox{\label{gbp:fig:pristPropsD}}{\includegraphics[scale=0.8]{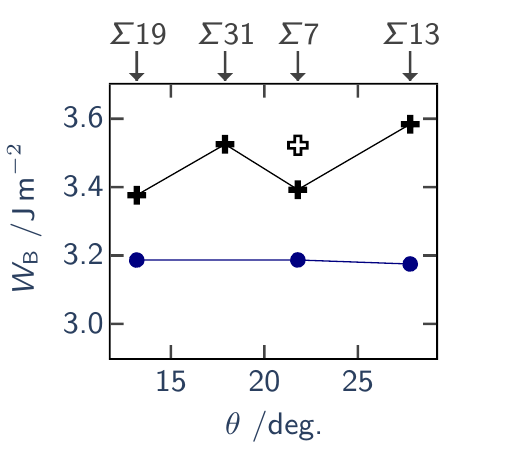}}%

    \caption{Interface energetics of five STGBs and three twist GBs: grain boundary energy \subref{gbp:fig:pristPropsA}, free surface energy \subref{gbp:fig:pristPropsB}, work of separation (GB) \subref{gbp:fig:pristPropsC}, and work of separation (bulk) \subref{gbp:fig:pristPropsD}.  Connecting lines are to guide the eye.}
    \label{gbp:fig:pristProps}
\end{figure}

Grain boundary energy, \en{GB}, free surface energy, \en{FS}, and work of ideal separation, \wSep{}{}, for each studied interface are listed in Table \ref{gbp:tab:pristineProps}. Fig.\ \ref{gbp:fig:pristProps} illustrates these quantities as a function of misorientation angle, \misori{}, and GB plane type (symmetric tilt, type I/II, or twist). No discernible correlation of interfacial energies with \misori{} is demonstrated, as was expected. The choice of boundary plane has a significant impact on these quantities. The twist GBs and FSs are relatively low energy interfaces. Consequently, the work of separation of the twist GBs is relatively high. We see the twist angle has a minimal impact on the interfacial energetics; all three twist interfaces exhibit $\en{GB} \approx \jpermsq{0.28}$, $\en{FS} \approx \jpermsq{1.59}$ and $\wSep{GB}{} \approx \jpermsq{2.90}$. We attribute the relatively low \en{GB} and high \wSep{GB}{} of the twist GBs to the similarity of the atomistic structures found at the twist GBs to the atomistic structure of bulk Zr. Variation in the twist angle does not substantially alter the structure.

\begin{table}[ht]
    \centering
    \caption{Properties of eight grain boundaries, including misorientation angle, \misori{}, and their corresponding free surfaces and bulk systems studied in this work. All GBs were formed by rotations about \hkl[0001]. Energies are expressed in \jpermsq{}.}
    \label{gbp:tab:pristineProps}
    \begin{tabular}{rlllllll}
        \toprule
        \sig{} & Plane              & Type    & \misori{}      & \en{GB} & \en{FS} & \wSep{GB}{} & \wSep{B}{} \\ \midrule
        7      & \gbPlane{s7}{tlA}  & Tilt I  & \gbRotAng{s7}  & 0.94    & 1.70    & 2.46        & 3.39       \\
        13     & \gbPlane{s13}{tlA} & Tilt I  & \gbRotAng{s13} & 0.51    & 1.79    & 3.08        & 3.58       \\
        19     & \gbPlane{s19}{tlA} & Tilt I  & \gbRotAng{s19} & 0.74    & 1.69    & 2.64        & 3.38       \\
        31     & \gbPlane{s31}{tlA} & Tilt I  & \gbRotAng{s31} & 0.71    & 1.76    & 2.81        & 3.53       \\
        7      & \gbPlane{s7}{tlB}  & Tilt II & \gbRotAng{s7}  & 0.69    & 1.76    & 2.83        & 3.52       \\
        7      & \gbPlane{s7}{tw}   & Twist   & \gbRotAng{s7}  & 0.28    & 1.59    & 2.91        & 3.19       \\
        13     & \gbPlane{s13}{tw}  & Twist   & \gbRotAng{s13} & 0.28    & 1.59    & 2.90        & 3.18       \\
        19     & \gbPlane{s19}{tw}  & Twist   & \gbRotAng{s19} & 0.29    & 1.59    & 2.90        & 3.18       \\ \bottomrule
    \end{tabular}
\end{table}

The relevance of the choice of GB plane in determining GB properties is most clearly demonstrated by the wide range of \en{GB} found for the \sig{7} GBs. Although the three \sig{7} GBs have an identical misorientation relationship, \en{GB} ranges from \jpermsq{0.28} for the twist GB, to \jpermsq{0.94} for the type I STGB, with the value of \en{GB} for the type II STGB between these, at \jpermsq{0.69}.

Generally, there is an inverse correlation between \en{GB} (Fig.\ \ref{gbp:fig:pristPropsA}) and \wSep{GB}{} (Fig.\ \ref{gbp:fig:pristPropsC}). For instance, the \sig{7} type I STGB has the highest \en{GB} of the studied systems, at \jpermsq{0.94}, but the lowest \wSep{GB}{}, at \jpermsq{2.46}. Intuitively, this indicates the least strongly-bound grain boundary has the lowest resistance to intergranular cleavage. However, it is important to note that \wSep{GB}{} is not a precise measure of GB strength, but rather an indicative measure, since it does not account for the energy barrier associated with cleavage.

In 2010, Janisch et al.\ \cite{Janischinitiotensiletests2010} identified (as part of an ab-initio tensile test study) the energetics of multiple GBs in face-centred cubic (FCC) \ce{Al}, including one twist GB: \sig{3}\hkl(111)\hkl[111]), and three STGBs: \sig{3}\hkl(11-2)\hkl[110], \sig{11}\hkl(1-13)\hkl[110], and \sig{9}\hkl(1-14)\hkl[110]. We note that, like our results, this study also found the twist GB to have the lowest \en{GB} (\jpermsq{0.048} for the twist GB versus between \jpermsq{0.171} and \jpermsq{0.486} for the STGBs). It would be interesting to know if the effect on \en{GB} of the twist angle for twist GBs is as small in FCC and/or body-centred cubic metals as we found it to be in hexagonal close-packed Zr. However, as far as we can tell, no other first-principles study has investigated the GB energy variation of twist GBs with twist angle.

In Fig.\ \ref{gbp:fig:pristPropsD}, we show the work of separation of bulk \alphaZr{} along the same cleavage planes as examined for each GB system. It can be seen that all bulk systems exhibit a larger work of separation compared to the GBs (i.e.\ $\wSep{B}{} > \wSep{GB}{}$). Relative to the bulk systems, the lack of translational symmetry at the GBs helps to explain this. With the exception of the \sig{13} STGB, the twist GBs have a larger work of separation than the STGBs. However, in the bulk systems, this ordering is reversed. In particular, the bulk cleavage planes corresponding to the STGBs have a larger work of separation.

\subsection{Interplanar spacing}

\begin{figure*}[ht]

    \hspace*{\fill}%
    \subcaptionbox{\sig{7} STGB I\label{gbp:fig:interplanarSpacingA}}{\includegraphics{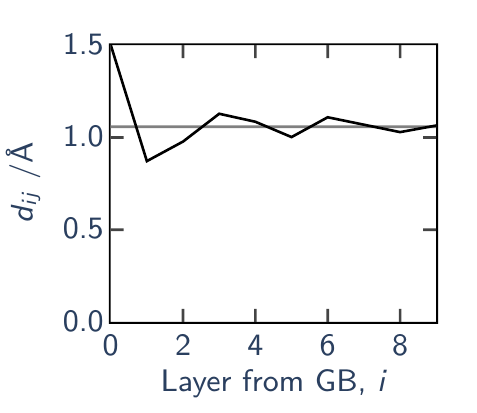}}%
    \subcaptionbox{\sig{7} STGB II\label{gbp:fig:interplanarSpacingB}}{\includegraphics{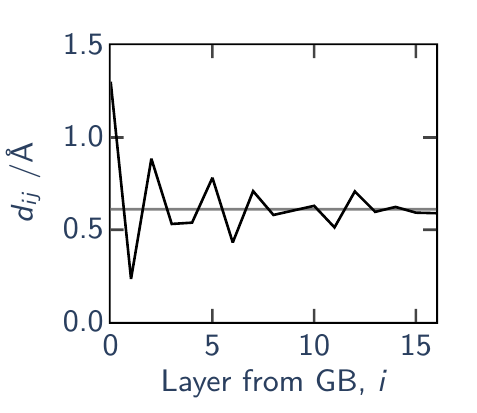}}%
    \subcaptionbox{\sig{13} STGB\label{gbp:fig:interplanarSpacingC}}{\includegraphics{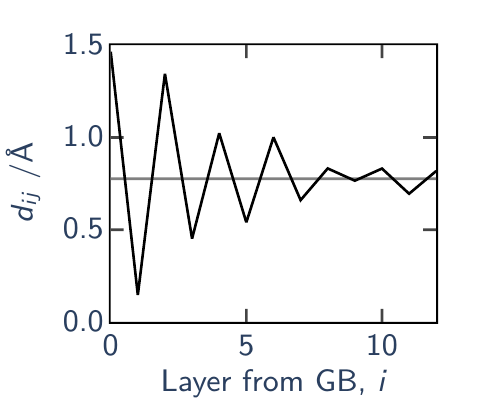}}%
    \hspace*{\fill}%

    \bigskip

    \hspace*{\fill}%
    \subcaptionbox{\sig{19} STGB\label{gbp:fig:interplanarSpacingD}}{\includegraphics{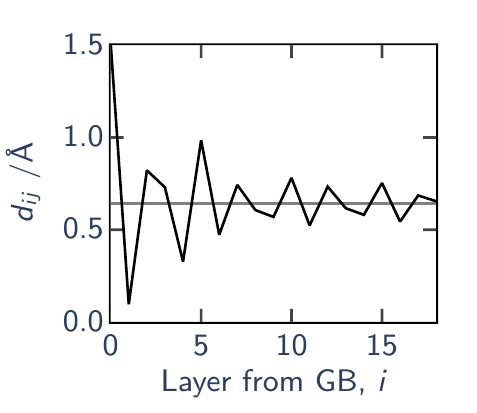}}%
    \subcaptionbox{\sig{31} STGB\label{gbp:fig:interplanarSpacingE}}{\includegraphics{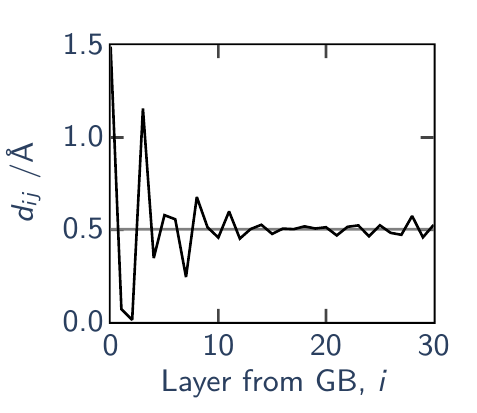}}%
    \subcaptionbox{Twist GBs\label{gbp:fig:interplanarSpacingF}}{\includegraphics{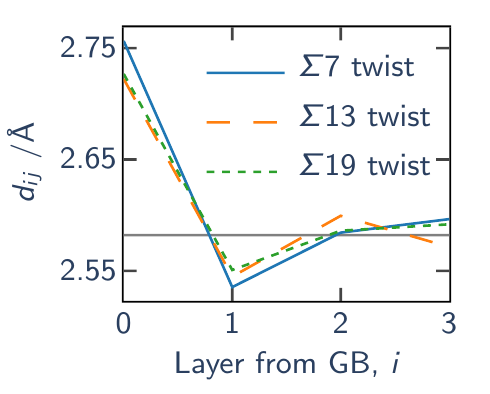}}%
    \hspace*{\fill}%

    \caption{Interplanar spacing of \subref{gbp:fig:interplanarSpacingA}--\subref{gbp:fig:interplanarSpacingE} the studied STGBs and \subref{gbp:fig:interplanarSpacingF} the studied twist GBs, recorded from the GB plane to the most bulk-like region of the supercell. The bulk interplanar spacing of each system is indicated with the horizontal line.}
    \label{gbp:fig:interplanarSpacing}
\end{figure*}

For each GB system, the spacing between adjacent crystallographic planes parallel to the GB plane is plotted as a function of plane number from the GB in Fig.\ \ref{gbp:fig:interplanarSpacing}. Results from the STGBs are shown in Figs.\ \ref{gbp:fig:interplanarSpacingA}--\subref{gbp:fig:interplanarSpacingD} and those for the twist GBs in Fig.\ \ref{gbp:fig:interplanarSpacingE}. In these figures, the $x$-axes extend from approximately the nominal GB plane at layer zero to the most bulk-like region of the supercell. In general, all GBs exhibit a common oscillatory response: an initial expansion relative to bulk interplanar spacing, followed by a contraction in the next layer.

For the STGBs, there is large variation in the magnitude of the changes in interplanar spacing. For example, the increase in interplanar spacing at the initial layer of the type I STGBs is found to increase with \sig{}-value (the bulk interplanar spacings are shown as horizontal lines in Fig.\ \ref{gbp:fig:interplanarSpacing}); the \sig{7} type I STGB exhibits an interplanar spacing increase at the initial layer of approximately \angs{0.4}, whereas the \sig{13}, \sig{19} and \sig{31} STGBs undergo expansions at the initial layers of approximately \angs{0.7}, \angs{0.9} and \angs{1.0}, respectively. This suggests an approximately common, optimal separation distance between the immediately adjacent layers in each micro-grain. In particular, the initial expansion of the first layers in the STGBs seem to be approximately \angs{1.5}. Furthermore, we see that the response of the \sig{13} STGB is more symmetric about the bulk interplanar spacing value. In other words, expansions and contractions in interplanar spacing are observed of approximately equal magnitude in this interface. In contrast, interplanar expansions are more prominent than contractions in the \sig{7}, \sig{19} and \sig{31} STGBs, and all the twist GBs.

For all the twist GBs, the interplanar spacings are very similar, and the maximum change is an expansion of approximately \angs{0.15}. In general, a change in interplanar spacing arises due to a lack of translational symmetry, resulting in interatomic bonding that is energetically less favourable than in bulk. The observed oscillatory response of the interplanar spacing illustrates how the initial expansion is accommodated by the remaining layers.

\subsection{Local atomic volume}

\begin{figure*}[ht]
    \hspace*{\fill}%
    \subcaptionbox{\label{gbp:fig:changeVolumePerAtomA}\sig{7} STGB I}{\includegraphics{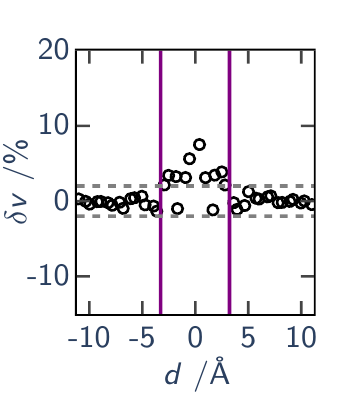}}%
    \subcaptionbox{\label{gbp:fig:changeVolumePerAtomB}\sig{7} STGB II}{\includegraphics{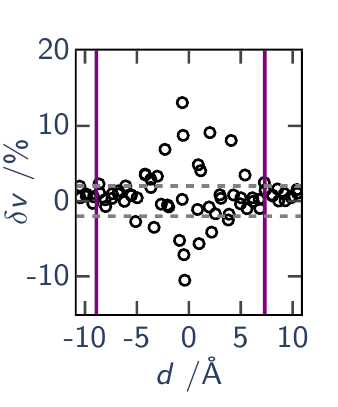}}%
    \subcaptionbox{\label{gbp:fig:changeVolumePerAtomC}\sig{13} STGB}{\includegraphics{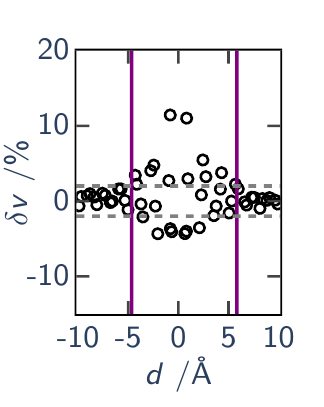}}%
    \subcaptionbox{\label{gbp:fig:changeVolumePerAtomD}\sig{19} STGB}{\includegraphics{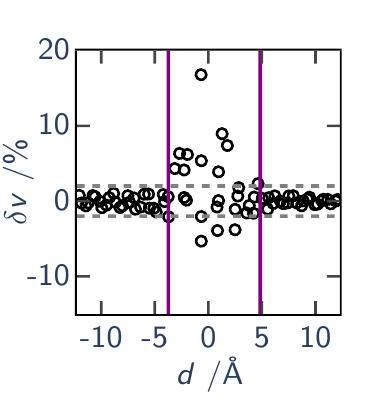}}%
    \hspace*{\fill}%

    \bigskip
    \hspace*{\fill}%
    \subcaptionbox{\label{gbp:fig:changeVolumePerAtomE}\sig{31} STGB}{\includegraphics{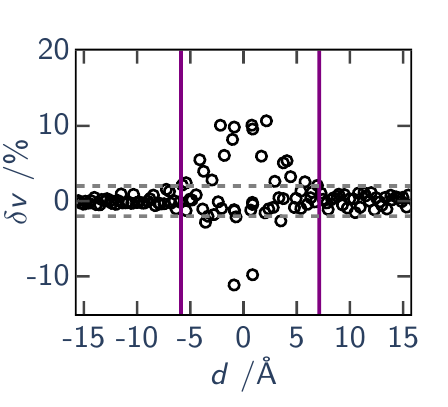}}%
    \subcaptionbox{\label{gbp:fig:changeVolumePerAtomF}\sig{7} twist GB}{\includegraphics{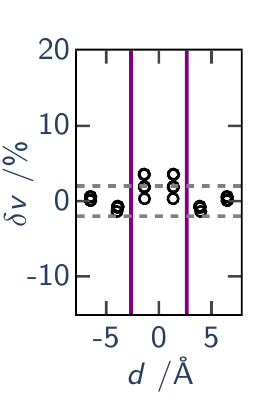}}%
    \subcaptionbox{\label{gbp:fig:changeVolumePerAtomG}\sig{13} twist GB}{\includegraphics{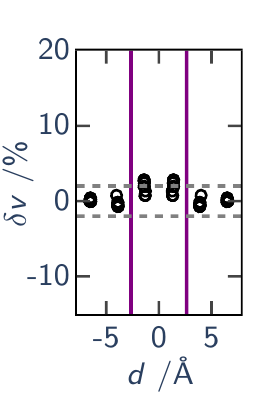}}%
    \subcaptionbox{\label{gbp:fig:changeVolumePerAtomH}\sig{19} twist GB}{\includegraphics{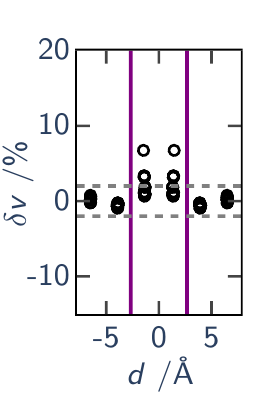}}%
    \hspace*{\fill}%

    \caption{Change in volume per atom relative to bulk \alphaZr{}, \atomVolChangeZr{}, as a function of distance from GB, \interfaceDist{}. An effective GB width, \gbWidth{}, has been calculated by considering the distances from the nominal GB plane at which $|\atomVolChangeZr{}| < \percent{2}$ for all remaining atoms towards the bulk-like regions. Vertical lines illustrate these distances, and dashed, grey, horizontal lines illustrate the threshold volume change used to calculate these distances.}
    \label{gbp:fig:changeVolumePerAtom}
\end{figure*}

By partitioning the supercell into regions associated with each atom, we elicit some measure of how the local atomic environment is modified due to the GB. We used a Voronoi tessellation to assign a convex polyhedron to each atom, and computed the volume of each region in the GB supercell after atomic relaxation. Fig.\ \ref{gbp:fig:changeVolumePerAtom} illustrates how the local atomic volume deviates from that of bulk \alphaZr{}, as a function of distance from the nominal GB plane. This deviation is represented as a percentage change in volume. Similarities can be observed between this data and the interplanar spacing data that we previously discussed. For instance, we note that in the vicinity of the \sig{13} STGB, some atoms undergo a local atomic volume contraction of nearly \percent{5}, whereas the largest atomic volume contraction in the \sig{7} type I STGB is around \percent{1}. Furthermore, it is evident that the distance over which atoms in the \sig{13} STGB experience a significant change in local volume is larger than that in the \sig{7} type I STGB. A similar pattern is observed in the response of the interplanar spacing in Fig.\ \ref{gbp:fig:interplanarSpacing}; the \sig{13} STGB exhibits a contraction of interplanar spacing that is much larger than that associated with the \sig{7} STGB.

The change in local atomic volume is largely similar across all of the twist GBs. Atomic regions in the plane nearest to the nominal GB plane experience an expansion, and those in the next-nearest plane experience a contraction of a smaller magnitude.

\subsection{GB width and maximum volume change}

Using the local atomic volume data portrayed in Fig.\ \ref{gbp:fig:changeVolumePerAtom}, we can quantify the relative thickness of the GB by considering a threshold deviation in local atomic volume. For example, we have included in Fig.\ \ref{gbp:fig:changeVolumePerAtom} horizontal dashed lines marking volume changes of \percent{\pm2}. We then derive the lower and upper limits of the spatial extent of the GB by considering the first atoms\,---\,as we consider atoms sequentially from the bulk region to the GB region\,---\,that experience a volume change larger than the specified threshold. Vertical solid lines are included in Fig.\ \ref{gbp:fig:changeVolumePerAtom} to indicated the width of the GB, \gbWidth{}, as determined in this way, where the lines are positioned at the midpoints between the final atom included by the threshold and the first atom excluded by the threshold.

It should be noted that the precise position of the vertical lines in Fig.\ \ref{gbp:fig:changeVolumePerAtom} that define \gbWidth{} are quite sensitive to the choice of volume change threshold. However, for the \percent{\pm2} threshold used here, the extents of the GBs defined in this way are approximately symmetric about the nominal GB plane.

\begin{figure}[ht]
    \subcaptionbox{GB width\label{gbp:fig:geomPropsWidth}}{\includegraphics[scale=0.8]{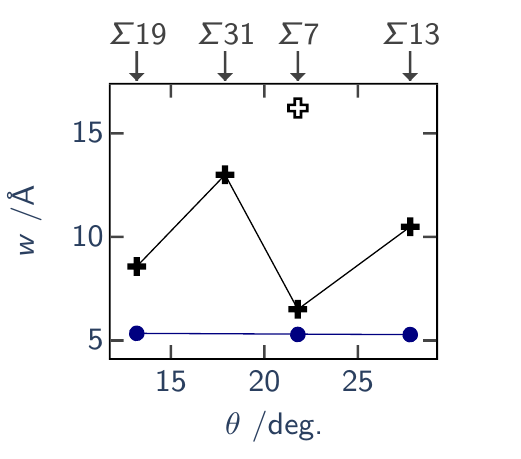}}%
    \subcaptionbox{GB expansion\label{gbp:fig:geomPropsExp}}{\includegraphics[scale=0.8]{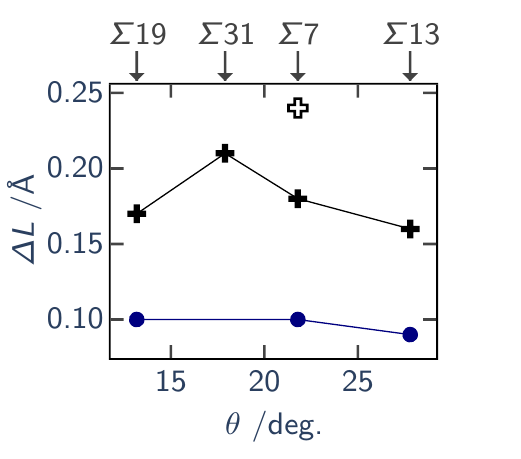}}%

    \bigskip
    \subcaptionbox{Maximum volume change\label{gbp:fig:geomPropsVolChange}}{\includegraphics[scale=0.8]{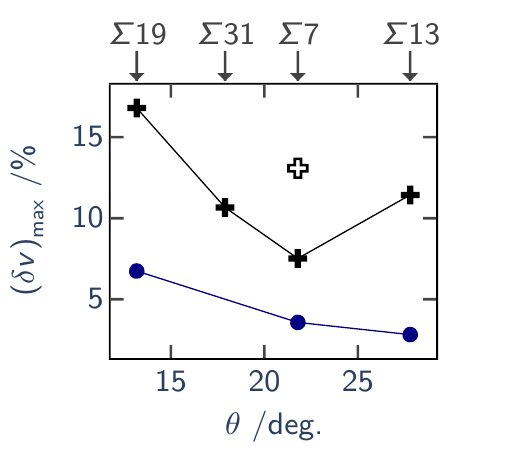}}%
    \subcaptionbox{GB energy\label{gbp:fig:geomPropsEnergy}}{\includegraphics[scale=0.8]{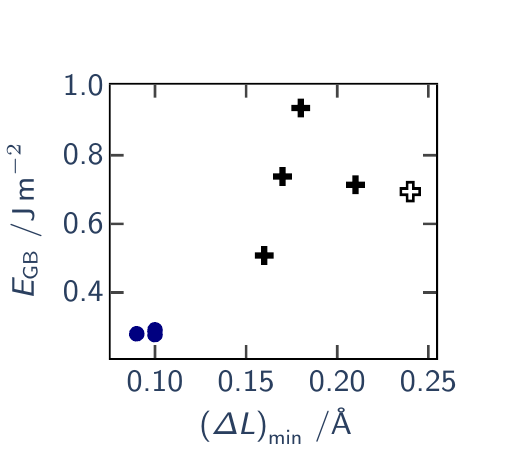}}%

    \vspace{0.1cm}%
    \hspace*{\fill}%
    \subcaptionbox*{}{\includegraphics[scale=0.8]{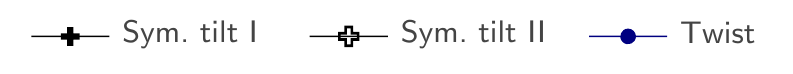}}%
    \hspace*{\fill}%
    \vspace{-0.8cm}%

    \caption{Geometric properties of the studied grain boundaries: \subref{gbp:fig:geomPropsWidth} the GB width defined by a threshold change in volume per atom of \percent{\pm{2}}; \subref{gbp:fig:geomPropsExp} the GB expansion of the supercell in the direction normal to the GB plane, resulting in a minimum-energy configuration; \subref{gbp:fig:geomPropsVolChange} the maximum volume per atom change relative to that of bulk \alphaZr{}; and \subref{gbp:fig:geomPropsEnergy} the relationship between minimum-energy GB expansion and grain boundary energy.}
    \label{gbp:fig:geomProps}
\end{figure}

The GB widths, \gbWidth{}, are plotted in Fig.\ \ref{gbp:fig:geomPropsWidth}, as a function of misorientation angle and GB plane type. It is interesting to compare \gbWidth{} with the minimum-energy GB expansions listed in Table \ref{gbp:tab:interfaceConstruction}. Thus, we include in Fig.\ \ref{gbp:fig:geomPropsExp} a comparable plot of this data. With perhaps the exception of the \sig{7} type I STGB, there is a correlation between the GB expansions and the GB widths of the STGBs. For instance, the \sig{7} type II STGB has both the largest GB expansion (\angs{0.24}) and the largest effective GB width (\angs{16}).

The twist GBs have the smallest effective widths, of around \angs{5}. The largest effective width, of around \angs{16}, is observed for the \sig{7} type II STGB. The response of the local atomic volume in the \sig{7} type II STGB is more similar to that in the \sig{31} STGB than the \sig{7} type I STGB. For instance, the effective widths are relatively large and there are significant contractions in the local atomic volumes of similar magnitudes in both the \sig{7} type II STGB and the \sig{31} STGB, whereas the \sig{7} type I STGB exhibits a smaller effective width and relatively small contractions in the local atomic volumes. This demonstrates how the \sig{}-value is not a good indicator of the local atomic environment, nor, by extension, the GB properties.

In Fig.\ \ref{gbp:fig:geomPropsVolChange}, we present the maximum increase in local atomic volume, over all supercell atoms, relative to the local atomic volume in bulk. The \sig{13} twist GB exhibits the smallest maximum change, of just \percent{2}; the largest maximum change is associated with the \sig{19} STGB, in which one atomic region experiences a volume change in excess of \percent{18}, relative to bulk.

Considering both GB width and maximum local volume change together as measures of the effect of a GB on the local atomic environment, we see the \sig{13} twist GB to have the smallest effect, with both a small width and a small maximum local atomic volume change. The \sig{19}, \sig{31} and \sig{7} type II STGBs can be seen to have the largest effect, with the largest effective widths and largest maximum changes in local atomic volume.

Furthermore, we also used our results to investigate the correlation between GB expansion, \gbExpMinSym{}, and GB energy, \en{GB}, as shown in Fig.\ \ref{gbp:fig:geomPropsEnergy}. We find a positive correlation; grain boundaries that preferentially adopt a larger GB expansion in order to accommodate their atomistic environment are associated with higher \en{GB}. This result agrees with that of Uesugi and Higashi (2011) \cite{UesugiFirstprinciplescalculationgrain2011}, who examined, from first principles, the nature of this relationship for six STGBs in \ce{Al}. This agreement is interesting, since in our case, the \sig{7} (type I), \sig{13} and \sig{31} STGBs had non-zero relative micro-grain translations in the $c$-direction, resulting in shifts between the basal planes of each of the micro-grains, whereas in the work of Uesugi and Higashi (2011), due to the cubic structure of \ce{Al}, the STGBs were more coherent.

\subsection{Atomic coordination}

\begin{figure*}[ht]
    \hspace*{\fill}%
    \subcaptionbox{\label{gbp:fig:coordinationA}\sig{7} STGB I}{\includegraphics{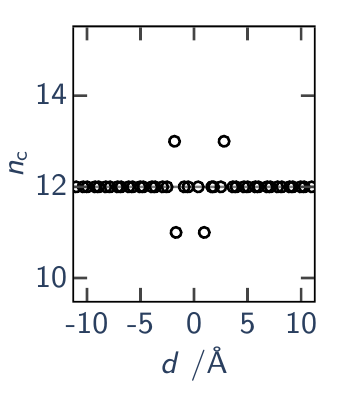}}%
    \subcaptionbox{\label{gbp:fig:coordinationB}\sig{7} STGB II}{\includegraphics{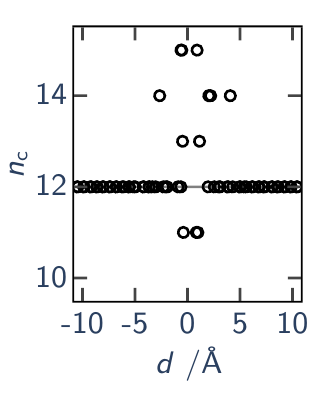}}%
    \subcaptionbox{\label{gbp:fig:coordinationC}\sig{13} STGB}{\includegraphics{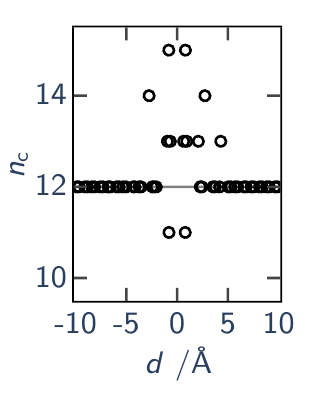}}%
    \subcaptionbox{\label{gbp:fig:coordinationD}\sig{19} STGB}{\includegraphics{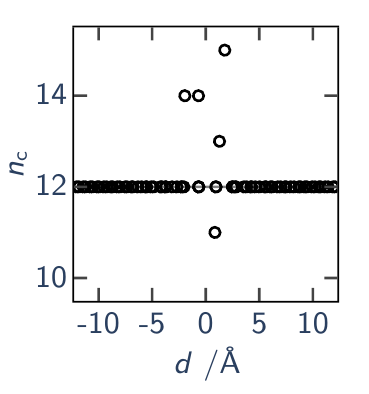}}%
    \hspace*{\fill}%

    \bigskip

    \hspace*{\fill}%
    \subcaptionbox{\label{gbp:fig:coordinationE}\sig{31} STGB}{\includegraphics{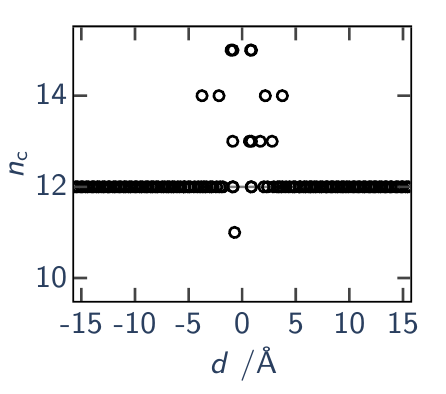}}%
    \subcaptionbox{\label{gbp:fig:coordinationF}\sig{7} twist GB}{\includegraphics{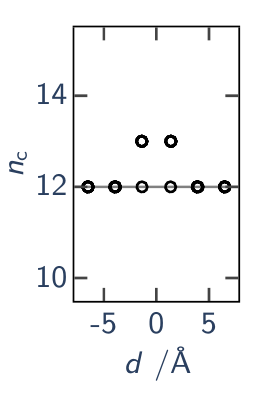}}%
    \subcaptionbox{\label{gbp:fig:coordinationG}\sig{13} twist GB}{\includegraphics{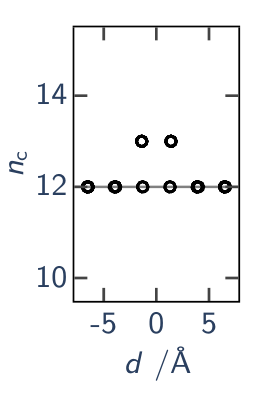}}%
    \subcaptionbox{\label{gbp:fig:coordinationH}\sig{19} twist GB}{\includegraphics{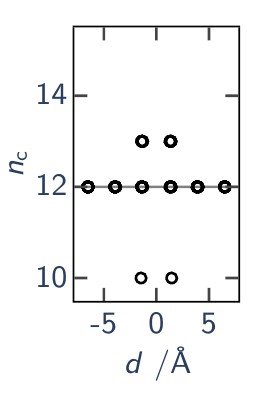}}%
    \hspace*{\fill}%
    \caption{Atomic coordination, \coordNum{}, determined by the number of facets on the Voronoi region of each atom, as a function of distance from the nominal GB plane, \interfaceDist{}. Facets smaller than \squareangs{1} are excluded. The coordination of bulk $\alpha$-Zr (12) is indicated as a horizontal line.}
    \label{gbp:fig:coordination}
\end{figure*}

In Fig.\ \ref{gbp:fig:coordination}, we illustrate how the local atomic coordination, \coordNum{}, changes with distance from the nominal GB plane. \coordNum{} was calculated from the Voronoi tessellation. Each Voronoi region associated with an atom has a number of facets, which are parallel to, and lie within, the bisection plane between a given atom and its neighbours. Simply counting the number of facets at each Voronoi region then gives a measure of the coordination. However, we also had to filter the results, such that facets with an area of less than \squareangs{1} were not counted. This was necessary, since periodic structures are associated with degenerate Voronoi vertices, which when numerically represented, will not be at precisely the same point. Thus extraneous facets are formed, which do not represent the presence of neighbours.

Our results show that, generally, due to the presence of a GB, atomic sites experience both increases and decreases in their coordination, which for bulk \alphaZr{} is $\coordNum{} = 12$. For all STGBs, except the \sig{7} type I STGB, several sites adopt high coordination numbers of up to $\coordNum{} = 15$, whereas only a few sites adopt lower-than-bulk coordination of $\coordNum{} = 11$. The \sig{7} type I STGB demonstrates with approximately equal frequency increases and decreases in coordination due to the GB. The \sig{19} twist GB is notable in that some atomic sites have coordination numbers of 10, two below the bulk value.

\subsection{Sharing of data}

During the course of this work, we developed a Python package named `atomistic' to assist with generating input structures and analysing simulation outputs \cite{Plowmanatomistic2020}. The code can additionally generate and analyse related \emph{sets} of structures, as is required when probing, for instance, a grain boundary's \gammaSurf{}.

Calculation of grain boundary structures can be time-consuming and computationally expensive, but is a necessary step in exploring the range of phenomena that occur at grain boundaries. To enable other researchers to make use of and build upon the results of our calculations, we have released our data on the public data repository Zenodo \cite{Plowmanaplowmanfirstprincipleszrgrainboundariesv12020a}. This repository, which points to a versioned GitHub repository, includes Python code that demonstrates our workflows. For example, we have included a text file that parametrises the structures in such a way that they can be programmatically reproduced using the `atomistic' code. Additionally, we have included the key CASTEP input and output files associated with our work. Detailed information about the contents of the repository can be found in the included `readme' file.

\section{Conclusions}

We conducted a first principles analysis on five symmetric tilt GBs and three twist GBs in Zr, generated using the first four low-\sig{} CSLs that have rotation axes parallel to the hexagonal $c$-axis of \alphaZr{}; STGBs and twist GBs were formed from \sig{7}, \sig{13} and \sig{19} CSLs, and the \sig{31} CSL was used to construct a further STGB. Using a plane-wave DFT code (CASTEP) we firstly explored the microscopic degrees of freedom of each boundary, in order to generate atomistic structures that were thermodynamically favourable. Generally, we found good agreement between DFT and an EAM empirical potential in predicting the energy landscape derived from sequential relative translations between the two micro-grains of each GB. Using GB, free surface and bulk models, we computed the interfacial energetics, such as grain boundary energy, \en{GB}, and work of ideal separation, \wSep{}{}, which provides an indicative measure of the resistance to decohesion. We found all three twist GBs to exhibit very similar interfacial energetics, whereas the STGBs demonstrated a larger variation. The choice of boundary plane (symmetric tilt or twist) was found to have a dramatic effect on the GB properties; for instance, the \sig{7} type I STGB has $\en{GB} = \jpermsq{0.94}$, whereas the \sig{7} twist GB has $\en{GB} = \jpermsq{0.28}$. The GB with the largest work of separation was the \sig{13} STGB, for which $\wSep{GB}{} = \jpermsq{3.08}$; whereas the smallest work of separation was found for the \sig{7} type I STGB, for which $\wSep{GB}{} = \jpermsq{2.46}$. In common with studies on other (cubic) metals \cite{ChenRolegrainboundary2017,BeanOrigindifferencesexcess2016,GuhlStructuralelectronicproperties2015,Janischinitiotensiletests2010}, we found that it is therefore important to consider a range of structures when studying phenomena that occur at grain boundaries; studying a single boundary is unlikely to provide representative results, especially if that boundary is an extreme case, such as the \sig{7} twist.

We additionally focussed on analysing the geometric properties of the GBs. For instance, we measured the effect of the GB on the interplanar spacing, between crystallographic planes that are parallel to the interface. A damped oscillatory response was observed. For most GBs, we saw an initial expansion of the interplanar spacing at the GB, followed by a contraction between the next-nearest planes. All twist GBs demonstrated an identical interplanar spacing response.

By computing the local volume per atom using a Voronoi tessellation, we were able to further examine the atomistic structures of the GBs. In particular, we considered how the volume of each atom changes due to the GB, and by setting a threshold volume change of \percent{\pm2}, were able to define an effective GB width, which ranged in length from \angs{5.28} for the \sig{13} twist GB, to \angs{16.22} for the \sig{7} type II STGB. Generally, we found the GB widths to correlate with the GB expansions associated with the minimum-energy structure. Furthermore, we used the Voronoi tessellation to measure how the local atomic coordination changed with distance from the GB. Coordination was found to vary from the bulk value of $\coordNum{}=12$ to between $\coordNum{}=10$ and $\coordNum{}=15$ for atom site near the GBs. However, typically, an increase in coordination was more prominent near the GBs. Finally, we also found a positive correlation between the minimum-energy GB expansion, and the GB energy: GBs that adopt a larger GB expansion tend to have a higher GB energy.

First principles analysis of GBs is valuable since it provides accurate information about the atomistic structures associated with GBs in real materials. Such information can assist in understanding, for instance, intergranular fracture mechanisms, such as those associated with the problem of the pellet-cladding interaction that we previously mentioned, since properties like the work of separation provide simple measures of resistance to decohesion. We will explore the segregation of impurities to these grain boundaries and the consequent embrittlement in a future publication \cite{PlowmanDensityFunctionalTheory2020}. Our present work could be built upon in several ways. We focussed primarily on examining the geometric properties of the studied systems (for instance, by investigating how the interplanar spacing, or local atomic volume respond to a GB). However, it would be informative to examine the electronic structure at the GBs. Doing so might assist in understanding the differences in the works of separation we observed for different GBs. Furthermore, it would also be interesting to explore more `type II' GB planes (i.e. additional tilt planes). In particular, we could then start to understand to what extent GB properties correlate with the choice of GB plane. For instance, can we measure structural differences between the types I/II as defined in this work, and does this result in distinct properties? Studying additional tilt boundaries would provide insight here.

\section*{Acknowledgements}
The authors gratefully acknowledge funding from the Engineering and Physical Sciences Research Council UK (EPSRC) and from Westinghouse Electric Sweden AB. C.\,P.\,Race was funded by a University Research Fellowship of the Royal Society, and the EPSRC MIDAS grant (EP/S01702X/1). A.\,J.\,Plowman was funded by the EPSRC PACIFIC grant (EP/L018616/1). Computational resources of the Computational Shared Facility at the University of Manchester and assistance provided by the Research IT team are also acknowledged.

\bibliography{gb_properties}

\end{document}